\documentclass[12pt]{article}
\usepackage{times}
\usepackage{natbib}
\usepackage{geometry}
\usepackage[utf8]{inputenc}
\usepackage{enumitem,amssymb}
\usepackage{ragged2e}
\usepackage[multiple]{footmisc}
\newlist{thematic}{itemize}{8}
\setlist[thematic]{label=$\square$}
\usepackage{pifont}
\newcommand{\cmark}{\ding{51}}%
\newcommand{\done}{\rlap{$\square$}{\raisebox{2pt}{\large\hspace{1pt}\cmark}}%
\hspace{-2.5pt}}

\newlength{\bibitemsep}
\newlength{\bibparskip}
\let\oldthebibliography\thebibliography
\renewcommand\thebibliography[1]{%
  \oldthebibliography{#1}%
  \setlength{\parskip}{\bibitemsep}%
  \setlength{\itemsep}{\bibparskip}%
}

\begin{document}

\begin{raggedright}  
\huge
Astro2020 Science White Paper \linebreak

The Role of Machine Learning in the Next Decade of Cosmology \linebreak
\normalsize

\noindent \textbf{Thematic Areas:} \hspace*{60pt} $\square$ Planetary Systems \hspace*{10pt} $\square$ Star and Planet Formation \hspace*{20pt}\linebreak
$\square$ Formation and Evolution of Compact Objects \hspace*{31pt} $\done$ Cosmology and Fundamental Physics \linebreak
  $\square$  Stars and Stellar Evolution \hspace*{1pt} $\square$ Resolved Stellar Populations and their Environments \hspace*{40pt} \linebreak
  $\square$    Galaxy Evolution   \hspace*{45pt} $\square$             Multi-Messenger Astronomy and Astrophysics \hspace*{65pt} \linebreak
  
\textbf{Principal Author:}

Name: Michelle Ntampaka
 \linebreak				
Institution: Harvard Data Science Initiative\\
Center for Astrophysics $|$ Harvard \& Smithsonian
 \linebreak
Email: michelle.ntampaka@cfa.harvard.edu
 \linebreak

\textbf{Co-authors:}
  Camille~Avestruz\footnotemark[1]\footnotemark[2],
  Steven~Boada\footnotemark[3],
  Jo\~ao~Caldeira\footnotemark[4],
  Jessi~Cisewski-Kehe\footnotemark[5],
  Rosanne~Di\thinspace Stefano\footnotemark[6],
  Cora~Dvorkin\footnotemark[7],
  August~E.~Evrard\footnotemark[8]\footnotemark[9],
  Arya~Farahi\footnotemark[10],
  Doug~Finkbeiner\footnotemark[6],
  Shy~Genel\footnotemark[11]\footnotemark[12],
  Alyssa~Goodman\footnotemark[6]\footnotemark[13]\footnotemark[14],
  Andy~Goulding\footnotemark[15],
  Shirley~Ho\footnotemark[10]\footnotemark[11]\footnotemark[15],
  Arthur~Kosowsky\footnotemark[16],
  Paul~La~Plante\footnotemark[17],
  Fran\c cois~Lanusse\footnotemark[18],
  Michelle~Lochner\footnotemark[19]\footnotemark[20],
  Rachel~Mandelbaum\footnotemark[10],
  Daisuke~Nagai\footnotemark[21],
  Jeffrey~A.~Newman\footnotemark[16],
  Brian~Nord\footnotemark[1]\footnotemark[4]\footnotemark[22],
  J.~E.~G.~Peek\footnotemark[23]\footnotemark[24],
  Austin~Peel\footnotemark[25],
  Barnab\'as~P\'oczos\footnotemark[26],
  Markus~Michael~Rau\footnotemark[10],
  Aneta~Siemiginowska\footnotemark[6],
  Danica~J.~Sutherland\footnotemark[27],
  Hy~Trac\footnotemark[10],
  Benjamin~Wandelt\footnotemark[11]\footnotemark[28]\footnotemark[29]
  \linebreak
  
\end{raggedright}
\begin{justify}
\noindent\textbf{Abstract:}\,\,\,\,\, In recent years, machine learning (ML) methods have remarkably improved how cosmologists can interpret data.  The next decade will bring new opportunities for data-driven cosmological discovery, but will also present new challenges for adopting ML methodologies and understanding the results.  ML could transform our field, but this transformation will require the astronomy community to both foster and promote interdisciplinary research endeavors.  
\end{justify}
\begin{raggedright}

\footnotetext[1]{\small{Kavli Institute for Cosmological Physics, The University of Chicago, Chicago, IL 60637, USA}}
\footnotetext[2]{\small{Enrico Fermi Institute, The University of Chicago, Chicago, IL 60637, USA}}
\footnotetext[3]{\small{Department of Physics and Astronomy, Rutgers, the State University of New Jersey, Piscataway, NJ 08854, USA}}
\footnotetext[4]{\small{Fermi National Accelerator Laboratory, Batavia, IL 60510, USA}}
\footnotetext[5]{\small{Department of Statistics and Data Science, Yale University, New Haven, CT 06511, USA}}
\footnotetext[6]{\small{Center for Astrophysics $|$ Harvard \& Smithsonian, Cambridge, MA 02138, USA}}
\footnotetext[7]{\small{Department of Physics, Harvard University, Cambridge, MA 02138, USA}}
\footnotetext[8]{\small{Department of Physics and Michigan Center for Theoretical Physics, University of Michigan, Ann Arbor, MI 48109, USA}}
\footnotetext[9]{\small{Department of Astronomy, University of Michigan, Ann Arbor, MI 48109, USA}}
\footnotetext[10]{\small{McWilliams Center for Cosmology, Department of Physics, Carnegie Mellon University, Pittsburgh, PA 15213, USA}}
\footnotetext[11]{\small{Center for Computational Astrophysics, Flatiron Institute, New York, NY 10010, USA}}
\footnotetext[12]{\small{Columbia Astrophysics Laboratory, Columbia University, New York, NY 10027, USA}}
\footnotetext[13]{\small{Harvard Data Science Initiative, Harvard University, Cambridge, MA 02138, USA}}
\footnotetext[14]{\small{Radcliffe Institute for Advanced Study, Harvard University, Cambridge, MA 02138, USA}}
\footnotetext[15]{\small{Department of Astrophysical Sciences, Princeton University, Princeton, 08544, USA}}
\footnotetext[16]{\small{Department of Physics and Astronomy, University of Pittsburgh, Pittsburgh, PA 15260, USA}}
\footnotetext[17]{\small{Department of Physics and Astronomy, University of Pennsylvania, Philadelphia, PA 19104, USA}}
\footnotetext[18]{\small{Berkeley Center for Cosmological Physics, University of California, Berkeley, CA 94720, USA}}
\footnotetext[19]{\small{African Institute for Mathematical Sciences, Cape Town, 7945, South Africa}}
\footnotetext[20]{\small{South African Radio Astronomy Observatory, Cape Town, 7405, South Africa}}
\footnotetext[21]{\small{Department of Physics, Yale University, New Haven, CT 06520, USA}}
\footnotetext[22]{\small{Department of Astronomy and Astrophysics, University of Chicago, Chicago, IL 60637, USA}}
\footnotetext[23]{\small{Space Telescope Science Institute, 3700 San Martin Dr, Baltimore, MD 21218, USA}}
\footnotetext[24]{\small{Department of Physics \& Astronomy, Johns Hopkins University, Baltimore, MD 21218, USA}}
\footnotetext[25]{\small{AIM, CEA, CNRS, Universit{\'e} Paris-Saclay, Universit{\'e} Paris Diderot, Sorbonne Paris Cit{\'e}, F-91191 Gif-sur-Yvette, France}}
\footnotetext[26]{\small{Machine Learning Department, Carnegie Mellon University, Pittsburgh, PA 15213, USA}}
\footnotetext[27]{\small{Gatsby Computational Neuroscience Unit, University College London, London, UK}}
\footnotetext[28]{\small{Sorbonne Universit\'e, CNRS, UMR 7095,  Institut d'Astrophysique de Paris (IAP), 75014 Paris, France}}
\footnotetext[29]{\small{Sorbonne Universit\'e, Institut Lagrange de Paris (ILP), 75014 Paris, France}}

\end{raggedright}

\clearpage

\begin{centering}
{\huge The Role of Machine Learning in the Next Decade of Cosmology}\\[5ex]
\end{centering}

\noindent Machine learning permeates our daily lives \textemdash{} performing tasks from  identifying the people in a photograph to  suggesting the next big purchase \textemdash{}  but will it change the way we do research?  The last decade has seen a remarkable rise in interdisciplinary machine learning (ML)-based astronomy research, offering enticing improvements in the ways we can interpret data. The next decade will see a continued rise in data-driven discovery as methods improve and data volumes grow, but realizing the full potential of ML \textemdash{} even in an era of unprecedented data volumes \textemdash{}  presents challenges.

\section{What Are Data Science and Machine Learning?}
\label{sec:datasci}

Data science is the study, application, and often the \textit{art} of creating and using sophisticated algorithms and cutting-edge data analysis techniques to extract information from data.  It includes the fields of statistics, machine learning, applied mathematics, and computer science.  Data science is an inherently interdisciplinary endeavor that reaches across departmental lines to produce new and innovative ways to interpret simulated data and astronomical observations.  When used well, data science provides methods for extracting more information from data sets than ever before, reducing bias and scatter, identifying interesting outliers, and inexpensively generating simulated data.  When used \textit{very} well, it guides our physical interpretation of observations and can lead to great discoveries.

While it's important to define what data science is, it's also important to define what it is \textit{not}. Data science is not the study of how to store or disseminate data.  While data storage and dissemination are important issues facing the astronomy community in the era of large surveys such as the Large Synoptic Survey Telescope (LSST), addressing these issues is not data science.  Nor is data science equivalent to data analysis.  When we refer to expanded uses of data science in cosmology, we do not include solidly established and easy-to-implement foundational tools such as chi-square analysis and linear regression.  Data science also is not ``big data,'' though the two  often work side-by-side.  

Data science encompasses a broad range of sophisticated data-analysis methodologies, and machine learning (ML) is just one tool in the toolbox.  Machine learning research explores the development and application of algorithms that find patterns in data.  In the context of astronomy, ML  algorithms can be used to address a broad range of tasks including:  describing complicated relationships, identifying data clusters and data outliers, reducing scatter by using complex or subtle signals, generating simulated data, classifying observations, addressing sparse data, and exploring data sets to understand the physical underpinning.  Machine learning is gaining traction within the astronomy community, and compelling successful applications of ML indicate that it has the potential to be transformative in the upcoming decade.

\section{Machine Learning Successes in Cosmology}
\label{sec:successes}

Recent successes illustrate the potential for sophisticated machine learning data-analysis tools to make significant strides in cosmology.  These successes include the following important results:

\begin{enumerate}[noitemsep]
	\item Galaxy clusters are sensitive to the underlying cosmological model, and low-scatter cluster mass proxies are one essential ingredient in using these objects to constrain parameters.  ML has been shown to significantly reduce scatter in cluster mass estimates compared to more traditional methods \citep{2015ApJ...803...50N, 2018arXiv181008430A, 2019arXiv190205950H}.  
	\item Weak lensing maps can shed light on the fundamental nature of gravity and cosmic acceleration. ML has been used with such maps to discriminate between standard and modified gravity models that generate statistically similar observations \citep{2018arXiv181011030P}.  Non-Gaussianities in weak lensing maps can encode cosmological information, but these are hard to measure or parameterize.  ML has been shown to tighten parameter constraints by a factor of five or more by harnessing these non-Gaussianities \citep{2018PhRvD..97j3515G, 2018arXiv180605995R}.  
	\item Next-generation cosmic microwave background (CMB) experiments will have increased sensitivity, enabling improved constraints on fundamental physics parameters. Achieving optimal constraints requires high signal-to-noise extraction of the projected gravitational potential from the CMB maps. ML techniques have been shown to provide competitive methods for this extraction, and are expected to excel in capturing hard-to-model non-Gaussian foreground and noise contributions~\citep{ 2018arXiv181001483C}.
	\item  $N$-body simulations are an effective approach to predicting structure formation of the universe, but are computationally expensive. ML has been used to predict structure formation of the universe, generating a full 3D $N$-body-like simulation with positions and velocities in $30ms$ \citep{2018arXiv181106533H}. This method outperforms traditional fast analytical approximation and accurately extrapolates far beyond its training data. 
    \item Estimating cosmological parameters from the large-scale structure is  traditionally done by calculating summary statistics of the observed large-scale structure traced by galaxies and then compared to the analytical theory. ML can be used to estimate cosmological parameters directly from the large-scale structure field and find more stringent constraints on the cosmological parameters \citep{2016arXiv160905796R}.
	\item Observations of the Epoch of Reionization can provide information about the earliest luminous sources.  ML can classify the types of sources driving reionization \citep{2018arXiv180703317H} and measure the duration of reionization to within 10\%, given a semi-analytic model and a strong prior on the midpoint of reionization \citep{2018arXiv181008211L}. 
	\item Topological data analysis (TDA) is an ML and statistical method for summarizing the shape of data.  TDA has been useful for discriminating dark energy models on simulated data \citep{van2011alpha}, isolating structures of the cosmic web \citep{sousbie2011persistent, libeskind2018tracing}, and defining new types of structures in the cosmic web such as filament loops \citep{xu2018finding}.  TDA may also help constrain the sum of neutrino masses \citep{xu2018finding}.
	\item Supernova classification is a critical step in obtaining cosmological constraints from type Ia supernovae in photometric surveys such as LSST. ML has proven to be a powerful tool \citep[e.g.,~][]{Lochner2016:1603.00882v3} and has been successfully applied to the current largest public supernova dataset \citep{Narayan2018:1801.07323v1}. The public has become heavily involved in developing new classification techniques \citep{team2018:1810.00001v1, Malz2018:1809.11145v1}.
	\item Strong lensing  probes cosmic structure along lines of sight.  ML was the most effective method at correctly identifying strong lensing arcs  in a recent data challenge, outperforming humans at this classification task \citep{2018MNRAS.473.3895L}. ML makes the analysis of strong lensing systems 10 million times faster than the state-of-the-art method \citep{2017Natur.548..555H, 2017ApJ...850L...7P}.

\end{enumerate}

\noindent ML will not displace standard statistical reasoning for well-modeled phenomena.  However, there are many cases where our current parametric models are inadequate to fully describe a physical system.  These ML successes  in cosmology imply that there is great potential for data-driven discovery, particularly as data sets grow and become more complex. 

\section{Challenges for the Next Decade}
\label{sec:challenges}

ML represents the next step in automation, driven by both rapidly increasing data volumes and the desire to prioritize human attention on tasks that require our insight and ingenuity.  There is no doubt that ML techniques will become more powerful and widespread in the 2020s, transforming our ability to address previously intractable problems.  

ML has demonstrated its potential to accelerate discovery in astrophysics, but challenges to more widespread adoption remain.  New tools come with new failure modes, and ML poses the temptation to choose expediency over understanding.  A common complaint about ML methods
is that they are black boxes that cannot lead to physical understanding, but this need not be the case.  Though it is difficult to understand the inner workings of a complex ML model, in many cases it is not impossible.  There are ways to peer inside the box and to gain physical understanding from complex models.  For example, Google's DeepDream project \citep{deepdream} originated as a way to visualize inputs that maximize activation in various layers of a neural network, and recent applications to cosmology indicate that the method can be used to gain physical understanding of astronomical systems \citep[e.g.,~][]{2018arXiv181007703N}.  Other recent developments to improve ML interpretability include saliency maps \citep{2013arXiv1312.6034S} that reveal which parts of an input most influence the output, and the deep k-nearest neighbors approach \citep{2018arXiv180304765P} that shows which training examples have the most influence on a specific outcome.  ML interpretability is an active area of research, and we  expect further improvements in the quality and diversity of  interpretation techniques in the next decade. 

At the intersection of ML and cosmology lies a unique opportunity for the benefit of both fields. ML is likely to accelerate discovery in cosmology through multiple applications and modalities (classification, regression, reinforcement learning). Cosmology, in return, provides new tasks and challenges for ML researchers and  well-understood data sets for testing ML methods.  The challenges provided by cosmology open opportunities for breakthroughs in the fundamental understanding of ML.

More advanced analyses of cosmological data place stringent requirements on the interpretability of results.  This represents a key hurdle for applying ML to cosmology: the assessment of  uncertainty and the removal of bias. Integrating traditional statistical methods with modern ML models may provide a solution, but this will require cross-disciplinary collaboration among statisticians, ML researchers, and cosmologists. During the 2020s, it is plausible that we could train, characterize, and use ML with the same rigor that we bring to more conventional statistical analysis. This is a significant shift in how we approach our research, and supporting this shift  will require the community's investment in education, interdisciplinary research endeavors, and the development and transfer of methodologies from the computer science community.  

\section{Opportunities in the 2020\MakeLowercase{s} and Beyond}
\label{sec:opportunities}

The assertion that ``astronomy is entering the era of big data'' has become clich\'e.  And yet, we cannot help but note that upcoming data sets, both big and small, will provide rich opportunities to use machine learning for teasing out complex correlations.  Here, we provide a few examples of those opportunities.\\

\textbf{Big Data Opportunities With LSST:}  The LSST survey \citep{2009arXiv0912.0201L} will provide the optical astronomical community with an unprecedented data rate.  It will cover nearly the entire visible southern sky roughly every three days for a decade, providing $\sim$1,000 exposures total at each position (split across 6 passbands).  With 500 petabytes of images, and a database including tens of trillions of observations of tens of billions of objects \citep{2008arXiv0805.2366I}, LSST's discovery potential will be enormous \textemdash{} but standard analysis methods will not enable the community to unlock the full potential of LSST. Its high source density, as well as the transient nature of some phenomena (e.g.,~asteroids and supernovae), will present a new set of challenges related to source identification and classification in this colossal dataset. For example, LSST expects to identify, and subsequently classify, 10 million rapid-response transient alerts on any given night. The continued development and future implementation of carefully designed ML algorithms at both the image processing \citep[e.g.,~][]{2018PASJ...70S..37G, 2018arXiv180710406D, 2018MNRAS.479..415A}
and catalog \citep[e.g.,~][]{Narayan2018:1801.07323v1, Malz2018:1809.11145v1} levels have the potential for producing significant advances in our ability to efficiently extract scientifically useful information (e.g.,~classification, distance, morphology, and mass) from the LSST data. However, these ML methodologies will require further exploration to fully understand their feasibility and general applicability to LSST.

\textbf{Big Data Opportunities in Radio Astronomy:}  The Hydrogen Epoch of Reionization Array (HERA), a radio interferometer seeking to provide the first detection of the 21 cm power spectrum from the Epoch of Reionization, is projected to produce 50 TB of data per night of observation when completed in late 2019. ML can identify radio frequency interference present in data from HERA faster and more reliably than traditional algorithms \citep{2019arXiv190208244K}. ML is also well-poised to replace other key aspects of the data analysis and reduction pipeline, such as the calibration of antennae and automatic identification of malfunctioning equipment. Further into the decade, the Square Kilometre Array (SKA) will feature data volumes even larger than HERA, with ML representing a viable path for analyzing and reducing these data in real time. 

\textbf{Pipeline Optimization Opportunities:}  ML has the potential to have a dramatic impact on the efficiency of cosmological experiments. For example, real-bogus \citep{2013MNRAS.435.1047B} is an ML system for determining whether a transient detected in photometric variation is a true variable object or simply an artifact. Similarly, the Dark Energy Camera Plane Survey \citep{2018ApJS..234...39S} uses a simple deep neural network to find images that have nebulosity, and, thus, require a separate processing algorithm. Furthermore, ML has the potential to simplify and accelerate the building of statistical inference pipelines in the context of full forward models through Likelihood-Free Inference \citep{2018MNRAS.477.2874A} and the automated extraction of informative features from data sets \citep{2018PhRvD..97h3004C}. The cosmological explorations of the 2020s will require excellent quality control while simultaneously handling unprecedented volumes of data, and ML is a strong option for pipeline optimization.

\textbf{Low Signal Opportunities With \textit{eROSITA}:}  While the most obvious targets of opportunity are observations with unprecedented data volumes, fully harnessing our smaller data sets provides rich opportunities for applying ML methodology as well.  For example, the upcoming \textit{eROSITA} mission is estimated to find more than 90,000 galaxy clusters with masses $> 10^{13.7}h^{-1}\,M_\odot$ \citep{2012MNRAS.422...44P}.  While the mission will detect clusters out to $z\sim 2$, a significant fraction of these cluster observations will be in the low-photon regime of 100 photons or fewer \citep{2018MNRAS.481..613P}.  Fully utilizing this sample will require developing techniques that provide low scatter mass proxies in the low signal regime, and ML is one viable option for this.

\textbf{Archival Data Opportunities:}  The potential for ML to help make great scientific strides is not limited to these upcoming data sets.  Research based on \textit{Hubble} archival data, for example, outnumbers those on new observations \citep{hubble}, showing that there is a vast, untapped potential even in the community's archival data.\\

The astronomy community already makes a significant investment in state-of-the-art instrumentation, software development to support data reduction \citep[e.g.,~][]{astropy:2013, 2015ascl.soft10007C, JHazelton2017, 2017ascl.soft03004M, astropy:2018}, and data management \citep[e.g.,~][]{2015AAS...22542204S, 2017AAS...22912801D}. There remains a strong need for this community to invest in the interdisciplinary development and application of cutting-edge ML techniques to interpret our rich and complex data, and help propel us into the next decade.

We encourage the astronomy community to invest in education and interdisciplinary research efforts that will transfer knowledge and methods from the ML research community to our field.  We also encourage efforts to build communities of practice for ML-based studies, especially those that profitably join simulated and observational survey data.  In the 2020s and beyond, these communities could cluster around discipline-focused hubs, or science gateways\footnote{https://sciencegateways.org/}, offering researchers access to open-source software, relevant data products, and common analysis workflows.

\section{Summary}
\label{sec:summary}

Recent applications of machine learning techniques to cosmological questions have made remarkable improvements in the way we interpret our data, and these compelling successes imply that ML has the potential to be transformative to our field.  
This transformation will require the astronomy community to cultivate and support research endeavors that cross traditional discipline boundaries, but the payoff has the potential to be steep. Machine learning will give cosmologists access to the data analysis methods that we need to fully utilize our rich data sets and make great scientific leaps forward over the next decade.

\clearpage
\section*{References}
\renewcommand\refname{}
\bibliography{../../papers/references}
\normalsize{\bibliographystyle{apj}}  	
\end{document}